\documentclass[twocolumn,superscriptaddress,nobalancelastpage,10pt,pra,letter]{revtex4-2}
\usepackage{amsmath}
\usepackage{amsthm}
\usepackage{amssymb}
\usepackage{amsfonts}
\usepackage{graphicx}
\usepackage{wasysym}
\usepackage{mathrsfs}
\usepackage{yfonts}
\usepackage{bbold}
\usepackage{verbatim}
\usepackage{subfigure}
\usepackage{color}
\usepackage{setspace}
\usepackage{multirow}
\usepackage{blkarray}

\makeatletter
\let\save@mathaccent\mathaccent
\newcommand*\if@single[3]{%
	\setbox0\hbox{${\mathaccent"0362{#1}}^H$}%
	\setbox2\hbox{${\mathaccent"0362{\kern0pt#1}}^H$}%
	\ifdim\ht0=\ht2 #3\else #2\fi
}
\newcommand*\rel@kern[1]{\kern#1\dimexpr\macc@kerna}
\newcommand*\widebar[1]{\@ifnextchar^{{\wide@bar{#1}{0}}}{\wide@bar{#1}{1}}}
\newcommand*\wide@bar[2]{\if@single{#1}{\wide@bar@{#1}{#2}{1}}{\wide@bar@{#1}{#2}{2}}}
\newcommand*\wide@bar@[3]{%
	\begingroup
	\def\mathaccent##1##2{%
		\let\mathaccent\save@mathaccent
		\if#32 \let\macc@nucleus\first@char \fi
		\setbox\z@\hbox{$\macc@style{\macc@nucleus}_{}$}%
		\setbox\tw@\hbox{$\macc@style{\macc@nucleus}{}_{}$}%
		\dimen@\wd\tw@
		\advance\dimen@-\wd\z@
		\divide\dimen@ 3
		\@tempdima\wd\tw@
		\advance\@tempdima-\scriptspace
		\divide\@tempdima 10
		\advance\dimen@-\@tempdima
		\ifdim\dimen@>\z@ \dimen@0pt\fi
		\rel@kern{0.6}\kern-\dimen@
		\if#31
		\overline{\rel@kern{-0.6}\kern\dimen@\macc@nucleus\rel@kern{0.4}\kern\dimen@}%
		\advance\dimen@0.4\dimexpr\macc@kerna
		\let\final@kern#2%
		\ifdim\dimen@<\z@ \let\final@kern1\fi
		\if\final@kern1 \kern-\dimen@\fi
		\else
		\overline{\rel@kern{-0.6}\kern\dimen@#1}%
		\fi
	}%
	\macc@depth\@ne
	\let\math@bgroup\@empty \let\math@egroup\macc@set@skewchar
	\mathsurround\z@ \frozen@everymath{\mathgroup\macc@group\relax}%
	\macc@set@skewchar\relax
	\let\mathaccentV\macc@nested@a
	\if#31
	\macc@nested@a\relax111{#1}%
	\else
	\def\gobble@till@marker##1\endmarker{}%
	\futurelet\first@char\gobble@till@marker#1\endmarker
	\ifcat\noexpand\first@char A\else
	\def\first@char{}%
	\fi
	\macc@nested@a\relax111{\first@char}%
	\fi
	\endgroup
}
\makeatother

\makeatletter
\newcommand*{\rom}[1]{\expandafter\@slowromancap\romannumeral #1@}
\makeatother

\makeatletter
\newcommand{\roml}[1]{\lowercase\expandafter{\romannumeral #1\relax}}
\makeatother

\DeclareMathAlphabet{\mathpzc}{OT1}{pzc}{m}{it}

\newcommand{\ket}[1]{| #1 \rangle}
\newcommand{\bra}[1]{\langle #1 |}

\newcommand{\V}{\mathcal{V}}

\begin{document}
	\title{High-dimensional quantum process tomography with undetected photons}
	
	\author{Salini Rajeev}
	\affiliation{Department of Physics, 145 Physical Sciences Bldg., Oklahoma State University, Stillwater, OK 74078, USA.}
	
	\author{Mayukh Lahiri}
	\email{mlahiri@okstate.edu} \affiliation{Department of Physics, 145 Physical Sciences Bldg., Oklahoma State University, Stillwater, OK 74078, USA.}
	
	
	\begin{abstract} 
		The goal of quantum process tomography is to fully characterize an operation performed on a quantum state. By considering high-dimensional quantum states (qudit), we show that it is possible to fully reconstruct an arbitrary operation without performing any measurement on the transformed qudit. Our method is interferometric and conceptually different from existing techniques of quantum process tomography that must perform a measurement on the transformed qudit. 
	\end{abstract}
	
	\maketitle
	Quantum process tomography (QPT) \textemdash the complete characterization of a quantum process \cite{chuang1997prescription,poyatos1997complete} \textemdash is a fundamental problem in quantum mechanics. QPT is also indispensable for quantum information science because accurate characterization of quantum circuit operations is essential for benchmarking, validating, and testing the performance of quantum networks \cite{hashim2025practical}. Although two-dimensional qubit systems provide valuable resources for quantum information processing, high-dimensional systems (i.e., qudits) turn out to be more desirable because they provide higher capacity of encoding and processing information \cite{erhard2020advances,chi2022programmable}. Furthermore, in communication with photonic states, qudit systems provide better key rates, improved security, and higher noise tolerance \cite{da2021path,islam2017provably,sheridan2010security}. 
	\par
	We consider the problem of completely characterizing any linear operation performed on a photonic qudit. All existing methods to perform this task rely on detecting the photon upon which the operation has acted (see, for example, \cite{bouchard2019quantum,mohseni2008quantum,escandon2024estimation}), even when the method is ancilla-assisted \cite{altepeter2003ancilla}. However, adequate single-photon detectors are not available for a wide spectral range. This fact poses a challenge to exploring various wavelength regions, for example, mid- and far-infrared regions. 
	\par
	It has been recently shown that quantum interference by path identity \cite{zou1991induced,hochrainer2022quantum} can be applied to fully reconstruct a high-dimensional unitary transformation without detecting the transformed qudit \cite{rajeev2026characterizing}. However, all quantum processes are not unitary. Non-unitary operations are often encountered in
	practical situations and demand separate attention. Here, we show that quantum interference by path identity can be applied to fully characterize any linear quantum operation that is not necessarily unitary. To demonstrate our method, we work with orbital angular momentum (OAM) states of light, which serve as a standard testbed for photonic qudits.

	\par
	Before considering the most general type of linear operations, it would be instructive to review why the analytical treatment used for unitary transformations is not applicable to the non-unitary ones. Let us consider an $N$-dimensional photonic qudit state given by
	\begin{align}\label{ref-state}
		\ket{\psi_0}=\frac{1}{\sqrt{N}}\sum_{l=0}^{N-1}\ket{l}=\frac{1}{\sqrt{N}}\sum_{l=0}^{N-1}\hat{a}^\dag(l)\ket{\text{vac}},
	\end{align}
	where $\ket{l}$ represents a photon in the OAM mode $l$, $\hat{a}^\dag(l)$ is the corresponding creation operator, and $\ket{\text{vac}}$ represents the vacuum state. In the paraxial limit, the creation and annihilation operators corresponding to the OAM modes must obey the commutation relation \cite{calvo2006quantum,plick2013quantum,karimi2014radial}
	\begin{align}\label{comm-rel}
		\big[\hat{a}(l), \hat{a}^\dag(m) \big]=\delta_{lm},
	\end{align}
	where $\delta_{lm}$ represents the Kronecker delta. When a unitary transformation, $\widehat{U}$, acts on the photonic qudit [Eq.~(\ref{ref-state})], the positive frequency component of the transformed quantized field in OAM mode $l$ is represented by 
	\begin{align}\label{U-trans}
		\hat{b}(l)=\sum_{\gamma=0}^{N-1}U_{l\gamma}\hat{a}(\gamma),
	\end{align}
	where $U_{l\gamma}=\bra{l} \widehat{U} \ket{\gamma}$ is a matrix element of $\widehat{U}$ in the OAM-basis obeying the unitarity condition $\sum_{\gamma=0}^{N-1}U_{l\gamma}U^*_{m\gamma}= \delta_{lm}$. It can be readily checked that a unitary transformation preserves the field commutation relation [Eq.~\eqref{comm-rel}], that is, $[\hat{b}(l), \hat{b}^\dag(m)]=\delta_{lm}$, ensuring that $\hat{b}$ represents a legitimate quantum field operator. 
	\par
	However, this is not the case with non-unitary transformations ($\widehat{T}$), for which 
	\begin{align}\label{non-U-def}
		\widehat{T}^{\dag} \widehat{T} \neq \mathbb{1} \quad \Longleftrightarrow \quad \sum_{\gamma=0}^{N-1}T_{l\gamma}T^*_{m\gamma} \neq \delta_{lm}.
	\end{align}
	If one assumes that a non-unitary operator transforms a quantum field in the same way a unitary operator does, i.e., if $\hat{b}(l)=\sum_{\gamma=0}^{N-1}T_{l\gamma}\hat{a}(\gamma)$, one immediately finds using Eq.~\eqref{non-U-def} that $[\hat{b}(l), \hat{b}^\dag(m)] \neq \delta_{lm}$; that is, $\hat{b}$ no longer represents a legitimate quantum field operator. 
	\par
	This issue can be taken care of by introducing the vacuum field. If a generally non-unitary operator, $\widehat{T}$, transforms a photonic quantum field $\hat{a}$ to $\hat{b}$, let us consider writing
	\begin{align}\label{non-U-trans}
		\hat{b}(l)=\sum_{\gamma=0}^{N-1}\big(T_{l\gamma}\hat{a}(\gamma)+A_{l\gamma}\hat{a}_0(\gamma) \big),
	\end{align}
	where $A_{l\gamma}$ is a complex coefficient that can be defined as an element of an $N\times N$ matrix, $\widehat{A}$, and $\hat{a}_0(\gamma)$ represents the vacuum field corresponding to mode $\gamma$, which obeys the commutaion relation $[\hat{a}_0(l), \hat{a}_0^\dag(m)]=\delta_{lm}$.
	It now follows from Eq.~(\ref{non-U-trans}) that
	\begin{align}\label{ref-n-trans-comm-reln-1}
		[\hat{b}(l),\hat{b}^\dag(m)] =\{\widehat{T} \, \widehat{T}^\dag\}_{lm}+\{\widehat{A}\widehat{A}^\dag\}_{lm}.
	\end{align}
	In order $\hat{b}$ to be a legitimate quantum field operator, the right-hand side of Eq.~\eqref{ref-n-trans-comm-reln-1} must be equal to $\delta_{lm}$; that is, we have 
	\begin{align}\label{A-cond}
		\{\widehat{A}\widehat{A}^\dag\}_{lm}=\delta_{lm}-\{\widehat{T} \, \widehat{T}^\dag\}_{lm}.
	\end{align}
	Equations (\ref{non-U-trans}) and (\ref{A-cond}) can be used to analytically treat a linear non-unitary operation in our case. 
	\par
	We now show that Eqs.~(\ref{non-U-trans}) and (\ref{A-cond}) also apply to unitary operators. If $\widehat{T}$ is unitary, i.e., if $\{\widehat{T} \, \widehat{T}^\dag\}_{lm}=\delta_{lm}$, it follows from Eq.~\eqref{A-cond} that in this case $\{\widehat{A}\widehat{A}^\dag\}_{lm}=0$ $\forall$ $l,m$; this is only possible when all matrix elements of $\widehat{A}$ are zero. Consequently, for a unitary operator, Eq.~(\ref{non-U-trans}) reduces to Eq.~(\ref{U-trans}). Therefore, \emph{the treatment that follows applies to both non-unitary and unitary operators.}
	\par
\begin{figure}[htbp]
		\centering
		\includegraphics[width=\linewidth]{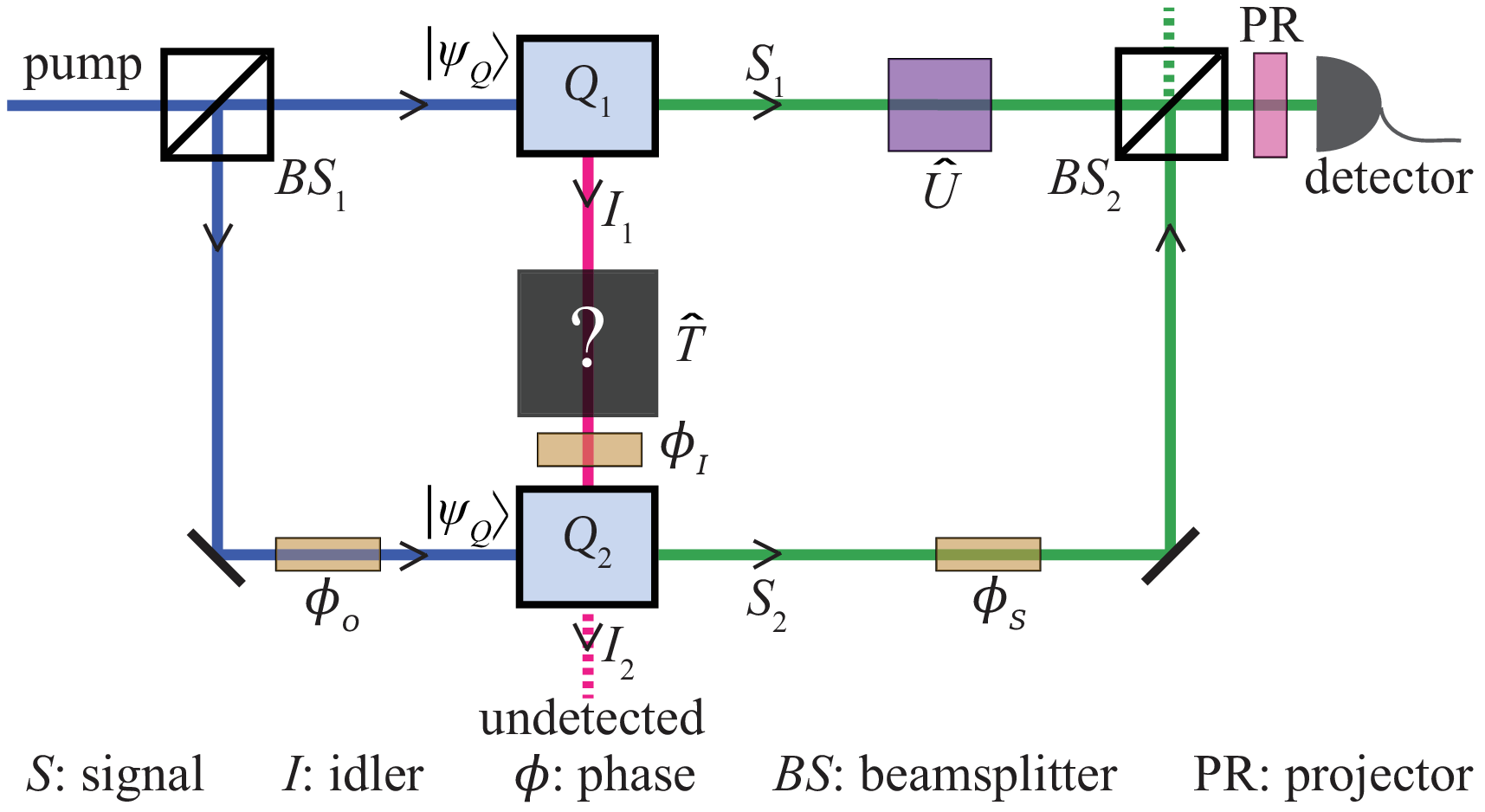}
		\caption{Quantum process tomography scheme. There are two identical sources, $Q_1$ and $Q_2$, individually emitting a signal-idler pair in a high-dimensional Bell state. Idler beam, $I_1$, is sent through $Q_2$ and perfectly aligned with idler beam $I_2$. This alignment induces coherence between signal beams $S_1$ and $S_2$. Consequently, a single-photon interference pattern is created when $S_1$ and $S_2$ are superposed by a beamsplitter ($BS_2$) and one of the outputs is detected after projecting the signal photon onto a chosen mode. An unknown high-dimensional operation ($\widehat{T}$) acts on the idler photon in beam $I_1$. Applying a suitable unitary transformation ($\widehat{U}$) in one of the signal beams, the unknown operation can be fully reconstructed from the interference pattern. The idler photon is never detected.}
		\label{fig:scheme}
	\end{figure}

	We now discuss the process tomography scheme (Fig.~\ref{fig:scheme}), which employs a Zou-Wang-Mandel interferometer \cite{zou1991induced}. There are two mutually coherent identical sources ($Q_1$ and $Q_2$), e.g., nonlinear crystals weakly pumped by mutually coherent laser beams. Each source can produce a photon pair constituted of an idler ($I$) and a signal ($S$) photon. Suppose that source $Q_j$ emits signal and idler photons into paraxial beams $S_j$ and $I_j$, respectively. Signal beams $S_1$ and $S_2$ are superposed by a beamsplitter. Photons emerging from one of the outputs of the beamsplitter is projected onto a suitable OAM state and then detected; only single-photon detection is performed. The idler beam ($I_1$) emerging from source $Q_1$ is sent through source $Q_2$ and then perfectly aligned with the idler beam ($I_2$) generated by $Q_2$. The most remarkable feature of the Zou-Wang-Mandel interferometer is that this alignment, which is often called path identity \cite{hochrainer2022quantum}, makes the which-path information unavailable for each signal photon \cite{zou1991induced}. Consequently, a single-photon interference pattern is observed at the detector. Note that the idler photon is not detected and no coincidence measurement or postselection is performed.
	\par
	Let us now consider a situation in which each source individually creates a high-dimensional Bell state, 
	\begin{align}\label{q-state-source}
		\ket{\psi_Q}=\frac{1}{\sqrt{N}}\sum_{k}^{N-1}\ket{k}_S\ket{f_k}_I =\frac{1}{\sqrt{N}}\sum_{k}^{N-1}\hat{a}^\dag_S(k)\hat{a}^\dag_I(f_k)\ket{\text{vac}},
	\end{align}
	where for a given $k$ there exists only one $f_k$. The unknown quantum operator $(\widehat{T})$, which is generally non-unitary, is applied to the idler photon between the two sources (Fig.~\ref{fig:scheme}). We also choose to apply a known and controllable unitary transformation ($\widehat{U}$) to the signal photon in one of the signal arms. We show below that information of both $\widehat{T}$ and $\widehat{U}$ appear in the single-photon interference pattern and that the unknown operator $\widehat{T}$ can be fully reconstructed by suitably choosing $\widehat{U}$. 
	\par
	As mentioned above, the two sources are weakly pumped by mutually coherent laser beams. In this case, the resulting quantum state is a linear superposition of the states produced by individual emissions \cite{zou1991induced}. In our case, using Eq.~(\ref{q-state-source}), we find the state to be given by
	\begin{align}\label{q-state-source-joint}
		\ket{\psi}&=\frac{1}{\sqrt{2N}}\sum_{k=0}^{N-1} (\ket{k}_{S_1}\ket{f_k}_{I_1}+e^{i\phi_0}\ket{k}_{S_2}\ket{f_k}_{I_2}),
	\end{align}
	where $1$ and $2$ in subscripts label the two sources, $\phi_0$ is a phase, and we have assumed for simplicity that the sources emit with the same probability. Note that the total photon occupation number of this state is two; multiple-pair generation and stimulated emission are negligible under the standard operating conditions of a Zou-Wang-Mandel interferometer as demonstrated by numerous experiments \cite{hochrainer2022quantum}. 
	\par
	When the generally non-unitary operator ($\widehat{T}$) acts on the idler field created by source $Q_1$, it transforms the field according to Eq.~(\ref{non-U-trans}). Therefore, the transformed field is given by
	\begin{align}
		\hat{b}_{I_1}(f_k)=\sum_{\gamma=0}^{N-1}\{ T_{k\gamma}\hat{a}_{I_1}(f_{\gamma})+A_{k\gamma} \, \hat{a}_{I_0}(f_\gamma)\},
	\end{align}
	where $\widehat{T}$ and $\widehat{A}$ are related by Eq.~\eqref{A-cond}. When path identity is applied, this transformed idler field must be the same as the idler field created at $Q_2$ except for a phase factor. We thus have 
	\begin{align}\label{align-cond-field}
		\hat{a}_{I_2}(f_k)&=e^{i\phi_I}\hat{b}_{I_1}(f_k) \nonumber\\&=e^{i\phi_I}\sum_{\gamma=0}^{N-1} \{ T_{k\gamma}\hat{a}_{I_1}(f_{\gamma})+A_{k\gamma} \hat{a}_{I_0}(f_\gamma) \},
	\end{align}
	where $\phi_I$ is the phase due to propagation from $Q_1$ to $Q_2$. Upon applying the Hermitian conjugate of Eq.~(\ref{align-cond-field}) to the vacuum state, we obtain
	\begin{align}\label{align-cond-kets}
		\ket{f_k}_{I_2}=e^{-i\phi_I}\sum_{\gamma=0}^{N-1}\left(T^*_{k\gamma}\ket{f_\gamma}_{I_1}+A_{k\gamma}^* \ket{f_\gamma}_{I_0}\right),
	\end{align}
	where $\ket{f_\gamma}_{I_0}=\hat{a}_{I_0}^{\dag}(f_\gamma) \ket{\text{vac}}$. Substituting for $\ket{f_k}_{I_2}$ from Eq.~\eqref{align-cond-kets} into Eq.~\eqref{q-state-source-joint}, we find that 
	\begin{align}\label{q-state-final}
		\ket{\psi}=&\frac{1}{\sqrt{2N}}\sum_{k=0}^{N-1}\Big[\ket{k}_{S_1}\ket{f_k}_{I_1}+e^{i(\phi_0-\phi_I)}\nonumber\\&\times\sum_{\gamma=0}^{N-1}\big(T^*_{k\gamma}\ket{k}_{S_2}\ket{f_\gamma}_{I_1}+A_{k\gamma}^*\ket{k}_{S_2}\ket{f_\gamma}_{I_0}\big)\Big].
	\end{align}
	Equation~\eqref{q-state-final} gives the quantum state generated in the interferometer before any transformation of the signal state is considered.
	\par
	When the known unitary transformation $\widehat{U}$ is applied to the signal photon in beam $S_1$ (Fig.~\ref{fig:scheme}), the corresponding quantum field transforms according to Eq.~\eqref{U-trans}; that is, we have
	\begin{align}\label{transformed-s1}
		\hat{b}_{S_1}(l)=\sum_{\gamma=0}^{N-1}U_{l\gamma} \, \hat{a}_{S_1}(\gamma).
	\end{align}
	After the application of $\widehat{U}$, the two signal beams are superposed by a beamsplitter and one of the outputs is sent to a single-photon detector (Fig.~\ref{fig:scheme}). Before arriving at the detector, the signal photon is projected onto a chosen OAM mode, say $d$. Therefore, the positive frequency part of the electric field operator at the detector is given by 
	\begin{align}\label{e-field-operator-0}
		\hat{E}_{d}^{(+)}\propto ie^{i\phi_S} \hat{a}_{S_2}(d) +\hat{b}_{S_1}(d),
	\end{align}
	where $\phi_S$ is the phase difference acquired due to propagation along beams $S_1$ and $S_2$. Using Eqs.~\eqref{transformed-s1} and \eqref{e-field-operator-0}, we obtain
	\begin{align}\label{e-field-operator}
		\hat{E}_{d}^{(+)}\propto ie^{i\phi_S} \hat{a}_{S_2}(d) +\sum_{\gamma=0}^{N-1}U_{d\gamma}\hat{a}_{S_1}(\gamma). 
	\end{align}
	We now apply the standard formula to determine the single-photon detection probability \cite{mandel1995optical}, which, in our case, is given by $P_d \propto \bra{\psi} \hat{E}_d^{(-)} \hat{E}_d^{(+)} \ket{\psi}$, where $\hat{E}_d^{(-)}=\{\hat{E}_d^{(+)} \}^{\dag}$ with $\dag$ representing Hermitian conjugation. Using Eqs.~\eqref{q-state-final} and \eqref{e-field-operator}, we readily find that 
	\begin{align}\label{int-pattern}
		P_d\propto 1+\sum_{\gamma=0}^{N-1}|U_{d\gamma}||T_{d\gamma}|\sin(\phi_{in}+\arg\{U_{d\gamma}\}+\arg\{T_{d\gamma}\}),
	\end{align}
	where $\phi_{in}=\phi_I-\phi_S-\phi_0$ is the tunable interferometric phase. Note that this single-photon detection probability is proportional to the single photon counting rate measured by the detector. One can observe in Eq.~\eqref{int-pattern} that $P_d$ varies sinusoidally with $\phi_{in}$; that is, $P_d$ represents a single-photon interference pattern. Furthermore, this interference pattern contains information about the matrix elements of both $\widehat{T}$ (unknown) and $\widehat{U}$ (known). We show below how to fully reconstruct $\widehat{T}$ by suitably choosing $\widehat{U}$. 
	\par
	We first consider the most trivial case in which the known unitary transformation is given by an identity matrix; that is, no transformation is applied to the signal photon before the beam splitter. In this case, we have $U_{d\gamma}=\delta_{d\gamma}$, where $\delta$ represents the Kronecker delta, and consequently, Eq.~\eqref{int-pattern} reduces to 
	\begin{align}\label{diag-elems}
		P_{d}\propto 1+ |T_{dd}|\sin(\phi_{in}+\arg\{T_{dd}\}).
	\end{align}
	It is evident that the complete information (magnitude and argument) of the diagonal elements of $\widehat{T}$ appears in these interference patterns. 
	\par
	We now target an arbitrary element of the unknown operator. Suppose that $\widehat{U}$ represents a two-dimensional rotation in the OAM-space; that is, it rotates two OAM-modes of the signal field, say $q$ and $r$, and leaves the rest unchanged. Let us denote such a matrix by $\widehat{U} \equiv \widehat{U}_b(q,r)$, whose elements are given by
	\begin{subequations}\label{u-k}
		\begin{align}
			&[\widehat{U}_b(q,r)]_{qq} = [\widehat{U}_b(q,r)]_{rr}= \cos\theta, \label{u-k:a} \\
			&[\widehat{U}_b(q,r)]_{qr} = -[\widehat{U}_b(q,r)]_{rq}=-\sin\theta , \label{u-k:b} \\
			&[\widehat{U}_b(q,r)]_{ll'} =
			\delta_{ll'},  ~\text{if}~ l\neq q,r ~ \text{and}~ l'\neq q,r, \label{u-k:c}
		\end{align}
	\end{subequations}
	where $\theta$ is the rotation angle in the OAM space and without any loss of generality we have assumed that $q<r$. We have chosen such a form of $\widehat{U}$ because this type of rotations are experimentally implementable \cite{babazadeh2017high, schlederer2016cyclic, brandt2020high, morizur2010programmable}. For our purpose, we need only two values of the rotation angle: $\theta=0$ and $\theta=\pi/2$. If we detect the modes that are subjected to rotation, we find using Eqs.~(\ref{int-pattern}) and (\ref{u-k}) that the corresponding single-photon detection probabilities are given by
	\begin{subequations} \label{p-d-int}
		\begin{align}
			&P^{q,r}_q(\theta=0)\propto 1+ |T_{qq}|\sin(\phi_{in}+\arg\{T_{qq}\}),\label{p-q-int-1} \\
			&P^{q,r}_q(\theta=\pi/2)\propto 1+ |T_{qr}|\sin(\phi_{in}+\pi+\arg\{T_{qr}\}),\label{p-q-int-2}\\
			&P^{q,r}_r(\theta=0)\propto 1+ |T_{rr}|\sin(\phi_{in}+\arg\{T_{rr}\}),\label{p-r-int-1}\\
			&P^{q,r}_r(\theta=\pi/2)\propto 1+ |T_{rq}|\sin(\phi_{in}+\arg\{T_{rq}\}), \label{p-r-int-2}
		\end{align}
	\end{subequations}
	where $q,r$ in the superscript implies that rotation was applied to signal modes $q$ and $r$. These four single-photon interference patterns contain complete information about the four matrix elements $T_{qq}$, $T_{qr}$, $T_{rq}$, and $T_{rr}$. Note that Eqs.~\eqref{p-q-int-1} and \eqref{p-r-int-1} yield the same result as Eq.~\eqref{diag-elems}; this is because when $\theta=0$, the known unitary transformation reduces to a unit matrix. 
	\par
\begin{figure*}
		\centering
		\includegraphics[width=\linewidth]{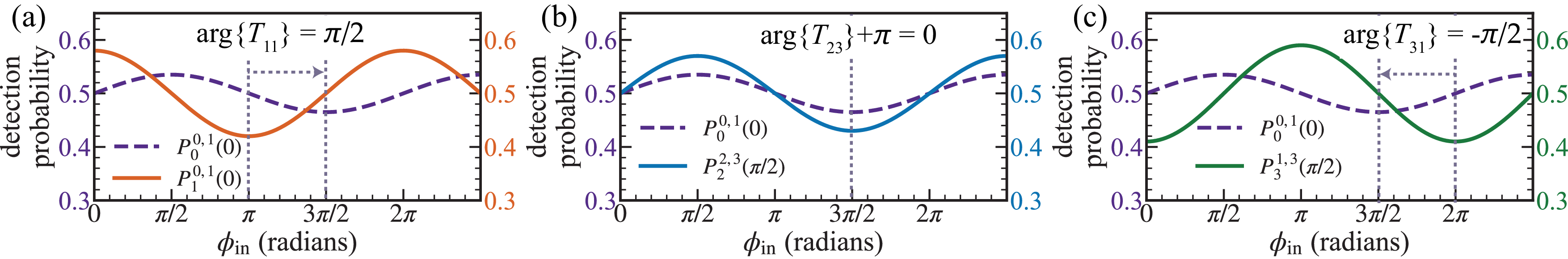}
		\caption{Reconstruction of a non-unitary operator is illustrated using the matrix elements $T_{11}$, $T_{23}$, and $T_{31}$. (a) The visibility of the interference pattern $P^{0,1}_1(\theta=0)$ gives $|T_{11}|=0.16$. The phase difference between $P^{0,1}_0(\theta=0)$ and $P^{0,1}_1(\theta=0)$ gives $\text{arg}\{T_{11}\}=\pi/2$. (b) The visibility of $P^{2,3}_2(\theta=\pi/2)$ gives $|T_{23}|=0.14$. A phase comparison between $P^{0,1}_0(\theta=0)$ and $P^{2,3}_2(\theta=\pi/2)$ yields $\text{arg}\{T_{23}\}+\pi=0$, which implies that $\text{arg}\{T_{23}\}=-\pi$. (c) The visibility of $P^{1,3}_3(\theta=\pi/2)$ gives $|T_{31}|=0.18$ and phase comparison with $P^{0,1}_0(\theta=0)$ shows that $\text{arg}\{T_{31}\}=-\pi/2$.}
		\label{fig:num-illust}
	\end{figure*}

	We now discuss how the matrix elements of the unknown operator ($\widehat{T}$) can be retrieved from the interference patterns. The visibility of the interference patterns can be determined by applying the standard formula
	\begin{align}\label{vis-def}
		\mathcal{V}_\mu (\theta)=\frac{[P_\mu (\theta)]_{\text{max}}-[P_\mu (\theta)]_{\text{min}}}{[P_\mu (\theta)]_{\text{max}}+[P_\mu (\theta)]_{\text{min}}},
	\end{align}
	where the maximum (max) and minimum (min) values of the single-photon detection probability are obtained by varying the interferometric phase $\phi_{in}$. Using Eqs.~(\ref{p-q-int-1})-(\ref{p-r-int-2}) and Eq.~\eqref{vis-def}, we find that 
	\begin{subequations}
		\begin{align}
			&\V^{q,r}_{q}(\theta=0)=|T_{qq}|, \hspace{0.5 cm} \V^{q,r}_{q}(\theta=\pi/2)=|T_{qr}|,\label{v-q}\\
			&\V^{q,r}_{r}(\theta=0)=|T_{rr}|, \hspace{0.5 cm} \V^{q,r}_{r}(\theta=\pi/2)=|T_{rq}|.\label{v-r}
		\end{align}
	\end{subequations}
	That is, the magnitudes of the matrix elements are given by the visibility of the interference patterns. The arguments of the matrix elements can be determined, for example, by comparing the phases of the interference patterns (illustrated in Fig.~\ref{fig:num-illust} using a numerical example). 
	\par
	We have thus shown that subjecting two signal modes ($q$ and $r$) to rotation in one of the signal beams and then detecting them after superposing the two signal beams allow one to retrieve complete information about four matrix elements ($T_{qq}$, $T_{qr}$, $T_{rq}$, and $T_{rr}$) of the unknown quantum operation. One can now vary $q$ and $r$ and retrieve all matrix elements of the unknown operation. Several protocols can be designed to achieve this. We discuss here the most intuitive protocol: (i) We first set $q=0$ and vary $r=1,2,\dots, N-1$; this allows to determine the following sets of matrix elements: $\{T_{00},T_{11}, \dots, T_{N-1 N-1} \}$, $\{T_{01},T_{02}, \dots T_{0N-1} \}$, $\{T_{10},T_{20}, \dots, T_{N-1 0} \}$. (ii) We then increase $q$ by steps of $1$ and in each step vary $r$ in the same manner as in (i) until we have retrieved all matrix elements. One can check that in order to reconstruct an $N$-dimensional unknown quantum operation in this manner, one needs $N(N-1)/2$ choices of $\widehat{U}_b(q,r)$ \cite{Note-comp-form}.
	\par 
	We now illustrate our results by considering a non-unitary operator, which takes the following form in the OAM-basis:  
	
	\begin{align}\label{had-loss}
		\widehat{T}=\begin{blockarray}{rrcccc}
			& & \ket{0} & \ket{1}  & \ket{2} & \ket{3}\\
			\begin{block}{rr[cccc]}
				\bra{0} &  & 0.07 & 0.07 & 0.07 & 0.07 \\
				\bra{1} & &  0.16 & 0.16i & -0.16&-0.16i\\
				\bra{2}& &0.14 &-0.14 &0.14 & -0.14\\
				\bra{3}&  &0.18 &-0.18i &-0.18 &0.18i\\
			\end{block}
		\end{blockarray}.  
	\end{align}
	
	This operator is obtained by introducing mode-dependent losses to the four-dimensional Hadamard gate. To fully reconstruct this matrix, we need $4\times(4-1)/2=6$ choices for the known unitary transformation: $\widehat{U}_b(0,1)$, $\widehat{U}_b(0,2)$, $\widehat{U}_b(0,3)$, $\widehat{U}_b(1,2)$, $\widehat{U}_b(1,3)$, and $\widehat{U}_b(2,3)$. Applying Eqs.~(\ref{p-q-int-1})-(\ref{p-r-int-2}), we obtain $16$ relevant single-photon interference patterns [Appendix A Eqs.~(\ref{p-h-a-0-1-had})-(\ref{p-h-f-3-had})], which are used to retrieve all $16$ matrix elements. In Fig.~\ref{fig:num-illust}, we illustrate the retrieval technique using three matrix elements: $T_{11}$, $T_{23}$, and $T_{31}$. Figure~\ref{fig:num-illust}a shows two interference patterns $P^{0,1}_0(\theta=0)$ and $P^{0,1}_1(\theta=0)$; their phase difference gives $\text{arg}\{T_{11}\}=\pi/2$. Similarly, in Figure~\ref{fig:num-illust}b, the phase difference between interference patterns $P^{0,1}_0(\theta=0)$ and $P^{2,3}_2(\theta=\pi/2)$ gives $\text{arg}\{T_{23}\}=-\pi$. Likewise, a comparison of $P^{0,1}_0(\theta=0)$ and $P^{1,3}_3(\theta=\pi/2)$ in Figure~\ref{fig:num-illust}c yields $\text{arg}\{T_{31}\}=-\pi/2$. The visibilities of the interference patterns $P^{0,1}_1(\theta=0)$, $P^{2,3}_2(\theta=\pi/2)$, and $P^{1,3}_3(\theta=\pi/2)$ give $|T_{11}|=0.16$, $|T_{23}|=0.14$, and $|T_{31}|=0.18$, respectively. Combining these magnitudes and arguments, we retrieve the three matrix elements: $T_{11}=0.16i$, $T_{23}=-0.14$, and $T_{31}=-0.18i$. Retrieval of the remaining matrix elements is given in Appendix A.
	\par
	In summary, we have proposed a method of quantum process tomography that does not require detecting the photon on which the unknown operation acted. Our results show that the path identity-based interferometric technique introduced to characterize unitary transformations can be applied to cover a much wider class of quantum operations. We have also shown how to treat high-dimensional non-unitary operations in path identity-based interferometers quantum mechanically. Finally, since our treatment is based on quantum field theory, it can, in principle, be extended to non-photonic quantum systems. 
	
	\section*{Acknowledgment} The research was supported by the Air Force Office of Scientific Research under grant FA9550-23-1-0216.
	\appendix
	\section{Calculations for the reconstruction of unknown operation}\label{app:A}
	The non-unitary operation chosen for numerical illustration is [Eq.~(\ref{had-loss}) in the main text]
	
	\begin{align}
		\widehat{T}=\begin{pmatrix}
			& 0.07 & 0.07 & 0.07 & 0.07\\
			&0.16 & 0.16i & -0.16&-0.16i\\
			&0.14 &-0.14 &0.14 & -0.14\\
			&0.18 &-0.18i &-0.18 &0.18i
		\end{pmatrix}.
	\end{align}
	
	We use Eqs.~(\ref{int-pattern}), (\ref{diag-elems}), (\ref{u-k}), and (\ref{p-d-int}) of the main text to obtain the following set of interference patterns corresponding to different choices of $\widehat{U}_b(q,r)$ (here we have chosen 1/2 as the proportionality constant):
	\begin{subequations}
		\begin{align}
			\text{for~}&\widehat{U}_b(0,1):\nonumber\\
			&P^{0,1}_{0} (\theta=0)=\frac{1}{2}\left[1+0.07\sin(\phi_{in})\right],\label{p-h-a-0-1-had}\\
			&P^{0,1}_{0} (\theta=\pi/2)=\frac{1}{2}\left[1-0.07\sin(\phi_{in})\right],\label{p-h-a-0-2-had}\\
			&P^{0,1}_{1}(\theta=0)=\frac{1}{2}\left[1+0.16\sin(\phi_{in}+\pi/2)\right]\label{p-h-a-1-1-had},\\
			&P^{0,1}_{1}(\theta=\pi/2)=\frac{1}{2}\left[1+0.16\sin(\phi_{in})\right]\label{p-h-a-1-2-had},\\
			\text{for~}&\widehat{U}_b(0,2):\nonumber\\
			&P^{0,2}_{0}(\theta=\pi/2)=\frac{1}{2}\left[1-0.07\sin(\phi_{in})\right],\label{p-h-b-0-1-had}\\
			&P^{0,2}_{2}(\theta=0)=\frac{1}{2}\left[1+0.14\sin(\phi_{in})\right]\label{p-h-b-2-1-had},\\
			&P^{0,2}_{2}(\theta=\pi/2)=\frac{1}{2}\left[1+0.14\sin(\phi_{in})\right]\label{p-h-b-2-2-had},\\
			\text{for~}&\widehat{U}_b(0,3):\nonumber\\
			&P^{0,3}_{0}(\theta=\pi/2)=\frac{1}{2}\left[1-0.07\sin(\phi_{in})\right],\label{p-h-c-0-1-had}\\
			&P^{0,3}_{3}(\theta=0)=\frac{1}{2}\left[1+0.18\sin(\phi_{in}+\pi/2)\right]\label{p-h-c-3-1-had},\\
			&P^{0,3}_{3}(\theta=\pi/2)=\frac{1}{2}\left[1+0.18\sin(\phi_{in})\right]\label{p-h-b-2-2-had},\\
			\text{for~}&\widehat{U}_b(1,2):\nonumber\\
			&P^{1,2}_{1}(\theta=\pi/2)=\frac{1}{2}\left[1-0.16\sin(\phi_{in}+\pi)\right],\label{p-h-d-1-had}\allowdisplaybreaks\\
			&P^{1,2}_{2}(\theta=\pi/2)=\frac{1}{2}\left[1+0.14\sin(\phi_{in}+\pi)\right]\label{p-h-d-2-had},\\
			\text{for~}&\widehat{U}_b(1,3):\nonumber\\
			&P^{1,3}_{1}(\theta=\pi/2)=\frac{1}{2}\left[1-0.16\sin(\phi_{in}-\pi/2)\right],\label{p-h-e-1-had}\\
			&P^{1,3}_{3}(\theta=\pi/2)=\frac{1}{2}\left[1+0.18\sin(\phi_{in}-\pi/2)\right]\label{p-h-d-2-had},\\
			\text{for~}&\widehat{U}_b(2,3):\nonumber\\
			&P^{2,3}_{2}(\theta=\pi/2)=\frac{1}{2}\left[1-0.14\sin(\phi_{in}+\pi)\right],\label{p-h-f-2-had}\\
			&P^{2,3}_{3}(\theta=\pi/2)=\frac{1}{2}\left[1+0.18\sin(\phi_{in}+\pi)\right]\label{p-h-f-3-had}.
		\end{align}
	\end{subequations}
	The magnitudes of the matrix elements are determined from the visibilities of the interference patterns (see Table.~\ref{tab-illustration-mag}).
	The arguments of the matrix elements are obtained from the phase difference between two suitably chosen interference patterns as explained in main text. We choose the interference pattern $P^{0,1}_0(\theta=0)$ as the reference and all the arguments are measured relative to this interference pattern (see Table.~\ref{tab-illustration-phase}). 
	\begin{table*}[htbp]
		\centering
		\caption{\bf Determination of $|T_{lm}|$ from single-photon interference patterns. }
		\begin{tabular}{cccc}
			\toprule
			magnitudes & choice of $\widehat{U}$ & visibility & values\\
			\hline
			$|T_{00}|$ & $\widehat{U}_b(0,1)$ & $\V^{0,1}_0(\theta=0)$ & 0.07\\
			$|T_{01}|$ & $\widehat{U}_b(0,1)$ & $\V^{0,1}_0(\theta=\pi/2)$ & 0.07\\
			$|T_{10}|$ & $\widehat{U}_b(0,1)$ & $\V^{0,1}_1(\theta=\pi/2)$ & 0.16\\
			$|T_{11}|$ & $\widehat{U}_b(0,1)$ & $\V^{0,1}_1(\theta=0)$ & 0.16\\
			$|T_{02}|$ & $\widehat{U}_b(0,2)$ & $\V^{0,2}_0(\theta=\pi/2)$ & 0.07\\
			$|T_{20}|$ & $\widehat{U}_b(0,2)$ & $\V^{0,2}_2(\theta=\pi/2)$ & 0.14\\
			$|T_{22}|$ & $\widehat{U}_b(0,2)$ & $\V^{0,2}_2(\theta=0)$ & 0.14\\
			$|T_{03}|$ & $\widehat{U}_b(0,3)$ & $\V^{0,3}_0(\theta=\pi/2)$ & 0.07\\
			$|T_{30}|$ & $\widehat{U}_b(0,3)$ & $\V^{0,3}_3(\theta=\pi/2)$ & 0.18\\
			$|T_{33}|$ & $\widehat{U}_b(0,3)$ & $\V^{0,3}_3(\theta=0)$ & 0.18\\
			$|T_{12}|$ & $\widehat{U}_b(1,2)$ & $\V^{1,2}_1(\theta=\pi/2)$ & 0.16\\
			$|T_{21}|$ & $\widehat{U}_b(1,2)$ & $\V^{1,2}_2(\theta=\pi/2)$ & 0.14\\
			$|T_{13}|$ & $\widehat{U}_b(1,3)$ & $\V^{1,3}_1(\theta=\pi/2)$ & 0.16\\
			$|T_{31}|$ & $\widehat{U}_b(1,3)$ & $\V^{1,3}_3(\theta=\pi/2)$ & 0.18\\
			$|T_{23}|$ & $\widehat{U}_b(2,3)$ & $\V^{2,3}_2(\theta=\pi/2)$ & 0.14\\
			$|T_{32}|$ & $\widehat{U}_b(2,3)$ & $\V^{2,3}_3(\theta=\pi/2)$ & 0.18\\
			\hline
		\end{tabular}
		\label{tab-illustration-mag}
	\end{table*}
	\begin{table*}[htbp]
		\centering
		\caption{\bf Determination of $\text{arg}\{T_{lm}\}$: the arguments are determined comparing the phase of an appropriate interference pattern with $P^{0,1}_0(\theta=0)$.}
		\begin{tabular}{cccc}
			\toprule
			arguments & choice of $\widehat{U}$ & interference pattern & values in radians\\
			\hline
			$\text{arg}\{T_{01}\}$ & $\widehat{U}_b(0,1)$ & $P^{0,1}_0(\theta=\pi/2)$ & 0\\
			$\text{arg}\{T_{10}\}$ & $\widehat{U}_b(0,1)$ & $P^{0,1}_1(\theta=\pi/2)$ & 0\\
			$\text{arg}\{T_{11}\}$ & $\widehat{U}_b(0,1)$ & $P^{0,1}_1(\theta=0)$ & $\pi/2$\\
			$\text{arg}\{T_{02}\}$ & $\widehat{U}_b(0,2)$ & $P^{0,2}_0(\theta=\pi/2)$ & 0\\
			$\text{arg}\{T_{20}\}$ & $\widehat{U}_b(0,2)$ & $P^{0,2}_2(\theta=\pi/2)$ & 0\\
			$\text{arg}\{T_{22}\}$ & $\widehat{U}_b(0,2)$ & $P^{0,2}_2(\theta=0)$ & 0\\
			$\text{arg}\{T_{03}\}$ & $\widehat{U}_b(0,3)$ & $P^{0,3}_0(\theta=\pi/2)$ & 0\\
			$\text{arg}\{T_{30}\}$ & $\widehat{U}_b(0,3)$ & $P^{0,3}_3(\theta=\pi/2)$ & 0\\
			$\text{arg}\{T_{33}\}$ & $\widehat{U}_b(0,3)$ & $P^{0,3}_3(\theta=0)$ & $\pi/2$\\
			$\text{arg}\{T_{12}\}$ & $\widehat{U}_b(1,2)$ & $P^{1,2}_1(\theta=\pi/2)$ & $\pi$\\
			$\text{arg}\{T_{21}\}$ & $\widehat{U}_b(1,2)$ & $P^{1,2}_2(\theta=\pi/2)$ & $\pi$\\
			$\text{arg}\{T_{13}\}$ & $\widehat{U}_b(1,3)$ & $P^{1,3}_1(\theta=\pi/2)$ & $-\pi/2$\\
			$\text{arg}\{T_{31}\}$ & $\widehat{U}_b(1,3)$ & $P^{1,3}_3(\theta=\pi/2)$ & $-\pi/2$\\
			$\text{arg}\{T_{23}\}$ & $\widehat{U}_b(2,3)$ & $P^{2,3}_2(\theta=\pi/2)$ & $\pi$\\
			$\text{arg}\{T_{32}\}$ & $\widehat{U}_b(2,3)$ & $P^{2,3}_3(\theta=\pi/2)$ & $\pi$\\
			\hline
		\end{tabular}
		\label{tab-illustration-phase}
	\end{table*}
	\clearpage
	\bibliography{ref}

\begin{thebibliography}{24}%
\makeatletter
\providecommand \@ifxundefined [1]{%
 \@ifx{#1\undefined}
}%
\providecommand \@ifnum [1]{%
 \ifnum #1\expandafter \@firstoftwo
 \else \expandafter \@secondoftwo
 \fi
}%
\providecommand \@ifx [1]{%
 \ifx #1\expandafter \@firstoftwo
 \else \expandafter \@secondoftwo
 \fi
}%
\providecommand \natexlab [1]{#1}%
\providecommand \enquote  [1]{``#1''}%
\providecommand \bibnamefont  [1]{#1}%
\providecommand \bibfnamefont [1]{#1}%
\providecommand \citenamefont [1]{#1}%
\providecommand \href@noop [0]{\@secondoftwo}%
\providecommand \href [0]{\begingroup \@sanitize@url \@href}%
\providecommand \@href[1]{\@@startlink{#1}\@@href}%
\providecommand \@@href[1]{\endgroup#1\@@endlink}%
\providecommand \@sanitize@url [0]{\catcode `\\12\catcode `\$12\catcode
  `\&12\catcode `\#12\catcode `\^12\catcode `\_12\catcode `\%12\relax}%
\providecommand \@@startlink[1]{}%
\providecommand \@@endlink[0]{}%
\providecommand \url  [0]{\begingroup\@sanitize@url \@url }%
\providecommand \@url [1]{\endgroup\@href {#1}{\urlprefix }}%
\providecommand \urlprefix  [0]{URL }%
\providecommand \Eprint [0]{\href }%
\providecommand \doibase [0]{https://doi.org/}%
\providecommand \selectlanguage [0]{\@gobble}%
\providecommand \bibinfo  [0]{\@secondoftwo}%
\providecommand \bibfield  [0]{\@secondoftwo}%
\providecommand \translation [1]{[#1]}%
\providecommand \BibitemOpen [0]{}%
\providecommand \bibitemStop [0]{}%
\providecommand \bibitemNoStop [0]{.\EOS\space}%
\providecommand \EOS [0]{\spacefactor3000\relax}%
\providecommand \BibitemShut  [1]{\csname bibitem#1\endcsname}%
\let\auto@bib@innerbib\@empty
\bibitem [{\citenamefont {Chuang}\ and\ \citenamefont
  {Nielsen}(1997)}]{chuang1997prescription}%
  \BibitemOpen
  \bibfield  {author} {\bibinfo {author} {\bibfnamefont {I.~L.}\ \bibnamefont
  {Chuang}}\ and\ \bibinfo {author} {\bibfnamefont {M.~A.}\ \bibnamefont
  {Nielsen}},\ }\bibfield  {title} {\bibinfo {title} {Prescription for
  experimental determination of the dynamics of a quantum black box},\
  }\href@noop {} {\bibfield  {journal} {\bibinfo  {journal} {Journal of Modern
  Optics}\ }\textbf {\bibinfo {volume} {44}},\ \bibinfo {pages} {2455}
  (\bibinfo {year} {1997})}\BibitemShut {NoStop}%
\bibitem [{\citenamefont {Poyatos}\ \emph {et~al.}(1997)\citenamefont
  {Poyatos}, \citenamefont {Cirac},\ and\ \citenamefont
  {Zoller}}]{poyatos1997complete}%
  \BibitemOpen
  \bibfield  {author} {\bibinfo {author} {\bibfnamefont {J.}~\bibnamefont
  {Poyatos}}, \bibinfo {author} {\bibfnamefont {J.~I.}\ \bibnamefont {Cirac}},\
  and\ \bibinfo {author} {\bibfnamefont {P.}~\bibnamefont {Zoller}},\
  }\bibfield  {title} {\bibinfo {title} {Complete characterization of a quantum
  process: the two-bit quantum gate},\ }\href@noop {} {\bibfield  {journal}
  {\bibinfo  {journal} {Physical Review Letters}\ }\textbf {\bibinfo {volume}
  {78}},\ \bibinfo {pages} {390} (\bibinfo {year} {1997})}\BibitemShut
  {NoStop}%
\bibitem [{\citenamefont {Hashim}\ \emph {et~al.}(2025)\citenamefont {Hashim},
  \citenamefont {Nguyen}, \citenamefont {Goss}, \citenamefont {Marinelli},
  \citenamefont {Naik}, \citenamefont {Chistolini}, \citenamefont {Hines},
  \citenamefont {Marceaux}, \citenamefont {Kim}, \citenamefont {Gokhale} \emph
  {et~al.}}]{hashim2025practical}%
  \BibitemOpen
  \bibfield  {author} {\bibinfo {author} {\bibfnamefont {A.}~\bibnamefont
  {Hashim}}, \bibinfo {author} {\bibfnamefont {L.~B.}\ \bibnamefont {Nguyen}},
  \bibinfo {author} {\bibfnamefont {N.}~\bibnamefont {Goss}}, \bibinfo {author}
  {\bibfnamefont {B.}~\bibnamefont {Marinelli}}, \bibinfo {author}
  {\bibfnamefont {R.~K.}\ \bibnamefont {Naik}}, \bibinfo {author}
  {\bibfnamefont {T.}~\bibnamefont {Chistolini}}, \bibinfo {author}
  {\bibfnamefont {J.}~\bibnamefont {Hines}}, \bibinfo {author} {\bibfnamefont
  {J.~P.}\ \bibnamefont {Marceaux}}, \bibinfo {author} {\bibfnamefont
  {Y.}~\bibnamefont {Kim}}, \bibinfo {author} {\bibfnamefont {P.}~\bibnamefont
  {Gokhale}}, \emph {et~al.},\ }\bibfield  {title} {\bibinfo {title} {Practical
  introduction to benchmarking and characterization of quantum computers},\
  }\href@noop {} {\bibfield  {journal} {\bibinfo  {journal} {PRX Quantum}\
  }\textbf {\bibinfo {volume} {6}},\ \bibinfo {pages} {030202} (\bibinfo {year}
  {2025})}\BibitemShut {NoStop}%
\bibitem [{\citenamefont {Erhard}\ \emph {et~al.}(2020)\citenamefont {Erhard},
  \citenamefont {Krenn},\ and\ \citenamefont {Zeilinger}}]{erhard2020advances}%
  \BibitemOpen
  \bibfield  {author} {\bibinfo {author} {\bibfnamefont {M.}~\bibnamefont
  {Erhard}}, \bibinfo {author} {\bibfnamefont {M.}~\bibnamefont {Krenn}},\ and\
  \bibinfo {author} {\bibfnamefont {A.}~\bibnamefont {Zeilinger}},\ }\bibfield
  {title} {\bibinfo {title} {Advances in high-dimensional quantum
  entanglement},\ }\href@noop {} {\bibfield  {journal} {\bibinfo  {journal}
  {Nature Reviews Physics}\ }\textbf {\bibinfo {volume} {2}},\ \bibinfo {pages}
  {365} (\bibinfo {year} {2020})}\BibitemShut {NoStop}%
\bibitem [{\citenamefont {Chi}\ \emph {et~al.}(2022)\citenamefont {Chi},
  \citenamefont {Huang}, \citenamefont {Zhang}, \citenamefont {Mao},
  \citenamefont {Zhou}, \citenamefont {Chen}, \citenamefont {Zhai},
  \citenamefont {Bao}, \citenamefont {Dai}, \citenamefont {Yuan} \emph
  {et~al.}}]{chi2022programmable}%
  \BibitemOpen
  \bibfield  {author} {\bibinfo {author} {\bibfnamefont {Y.}~\bibnamefont
  {Chi}}, \bibinfo {author} {\bibfnamefont {J.}~\bibnamefont {Huang}}, \bibinfo
  {author} {\bibfnamefont {Z.}~\bibnamefont {Zhang}}, \bibinfo {author}
  {\bibfnamefont {J.}~\bibnamefont {Mao}}, \bibinfo {author} {\bibfnamefont
  {Z.}~\bibnamefont {Zhou}}, \bibinfo {author} {\bibfnamefont {X.}~\bibnamefont
  {Chen}}, \bibinfo {author} {\bibfnamefont {C.}~\bibnamefont {Zhai}}, \bibinfo
  {author} {\bibfnamefont {J.}~\bibnamefont {Bao}}, \bibinfo {author}
  {\bibfnamefont {T.}~\bibnamefont {Dai}}, \bibinfo {author} {\bibfnamefont
  {H.}~\bibnamefont {Yuan}}, \emph {et~al.},\ }\bibfield  {title} {\bibinfo
  {title} {A programmable qudit-based quantum processor},\ }\href@noop {}
  {\bibfield  {journal} {\bibinfo  {journal} {Nature Communications}\ }\textbf
  {\bibinfo {volume} {13}},\ \bibinfo {pages} {1166} (\bibinfo {year}
  {2022})}\BibitemShut {NoStop}%
\bibitem [{\citenamefont {Da~Lio}\ \emph {et~al.}(2021)\citenamefont {Da~Lio},
  \citenamefont {Cozzolino}, \citenamefont {Biagi}, \citenamefont {Ding},
  \citenamefont {Rottwitt}, \citenamefont {Zavatta}, \citenamefont {Bacco},\
  and\ \citenamefont {Oxenl{\o}we}}]{da2021path}%
  \BibitemOpen
  \bibfield  {author} {\bibinfo {author} {\bibfnamefont {B.}~\bibnamefont
  {Da~Lio}}, \bibinfo {author} {\bibfnamefont {D.}~\bibnamefont {Cozzolino}},
  \bibinfo {author} {\bibfnamefont {N.}~\bibnamefont {Biagi}}, \bibinfo
  {author} {\bibfnamefont {Y.}~\bibnamefont {Ding}}, \bibinfo {author}
  {\bibfnamefont {K.}~\bibnamefont {Rottwitt}}, \bibinfo {author}
  {\bibfnamefont {A.}~\bibnamefont {Zavatta}}, \bibinfo {author} {\bibfnamefont
  {D.}~\bibnamefont {Bacco}},\ and\ \bibinfo {author} {\bibfnamefont {L.~K.}\
  \bibnamefont {Oxenl{\o}we}},\ }\bibfield  {title} {\bibinfo {title}
  {Path-encoded high-dimensional quantum communication over a 2-km multicore
  fiber},\ }\href@noop {} {\bibfield  {journal} {\bibinfo  {journal} {npj
  Quantum Information}\ }\textbf {\bibinfo {volume} {7}},\ \bibinfo {pages}
  {63} (\bibinfo {year} {2021})}\BibitemShut {NoStop}%
\bibitem [{\citenamefont {Islam}\ \emph {et~al.}(2017)\citenamefont {Islam},
  \citenamefont {Lim}, \citenamefont {Cahall}, \citenamefont {Kim},\ and\
  \citenamefont {Gauthier}}]{islam2017provably}%
  \BibitemOpen
  \bibfield  {author} {\bibinfo {author} {\bibfnamefont {N.~T.}\ \bibnamefont
  {Islam}}, \bibinfo {author} {\bibfnamefont {C.~C.~W.}\ \bibnamefont {Lim}},
  \bibinfo {author} {\bibfnamefont {C.}~\bibnamefont {Cahall}}, \bibinfo
  {author} {\bibfnamefont {J.}~\bibnamefont {Kim}},\ and\ \bibinfo {author}
  {\bibfnamefont {D.~J.}\ \bibnamefont {Gauthier}},\ }\bibfield  {title}
  {\bibinfo {title} {Provably secure and high-rate quantum key distribution
  with time-bin qudits},\ }\href@noop {} {\bibfield  {journal} {\bibinfo
  {journal} {Science Advances}\ }\textbf {\bibinfo {volume} {3}},\ \bibinfo
  {pages} {e1701491} (\bibinfo {year} {2017})}\BibitemShut {NoStop}%
\bibitem [{\citenamefont {Sheridan}\ and\ \citenamefont
  {Scarani}(2010)}]{sheridan2010security}%
  \BibitemOpen
  \bibfield  {author} {\bibinfo {author} {\bibfnamefont {L.}~\bibnamefont
  {Sheridan}}\ and\ \bibinfo {author} {\bibfnamefont {V.}~\bibnamefont
  {Scarani}},\ }\bibfield  {title} {\bibinfo {title} {Security proof for
  quantum key distribution using qudit systems},\ }\href@noop {} {\bibfield
  {journal} {\bibinfo  {journal} {Physical Review A}\ }\textbf {\bibinfo
  {volume} {82}},\ \bibinfo {pages} {030301} (\bibinfo {year}
  {2010})}\BibitemShut {NoStop}%
\bibitem [{\citenamefont {Bouchard}\ \emph {et~al.}(2019)\citenamefont
  {Bouchard}, \citenamefont {Hufnagel}, \citenamefont {Koutn{\`y}},
  \citenamefont {Abbas}, \citenamefont {Sit}, \citenamefont {Heshami},
  \citenamefont {Fickler},\ and\ \citenamefont {Karimi}}]{bouchard2019quantum}%
  \BibitemOpen
  \bibfield  {author} {\bibinfo {author} {\bibfnamefont {F.}~\bibnamefont
  {Bouchard}}, \bibinfo {author} {\bibfnamefont {F.}~\bibnamefont {Hufnagel}},
  \bibinfo {author} {\bibfnamefont {D.}~\bibnamefont {Koutn{\`y}}}, \bibinfo
  {author} {\bibfnamefont {A.}~\bibnamefont {Abbas}}, \bibinfo {author}
  {\bibfnamefont {A.}~\bibnamefont {Sit}}, \bibinfo {author} {\bibfnamefont
  {K.}~\bibnamefont {Heshami}}, \bibinfo {author} {\bibfnamefont
  {R.}~\bibnamefont {Fickler}},\ and\ \bibinfo {author} {\bibfnamefont
  {E.}~\bibnamefont {Karimi}},\ }\bibfield  {title} {\bibinfo {title} {Quantum
  process tomography of a high-dimensional quantum communication channel},\
  }\href@noop {} {\bibfield  {journal} {\bibinfo  {journal} {Quantum}\ }\textbf
  {\bibinfo {volume} {3}},\ \bibinfo {pages} {138} (\bibinfo {year}
  {2019})}\BibitemShut {NoStop}%
\bibitem [{\citenamefont {Mohseni}\ \emph {et~al.}(2008)\citenamefont
  {Mohseni}, \citenamefont {Rezakhani},\ and\ \citenamefont
  {Lidar}}]{mohseni2008quantum}%
  \BibitemOpen
  \bibfield  {author} {\bibinfo {author} {\bibfnamefont {M.}~\bibnamefont
  {Mohseni}}, \bibinfo {author} {\bibfnamefont {A.~T.}\ \bibnamefont
  {Rezakhani}},\ and\ \bibinfo {author} {\bibfnamefont {D.~A.}\ \bibnamefont
  {Lidar}},\ }\bibfield  {title} {\bibinfo {title} {Quantum-process tomography:
  Resource analysis of different strategies},\ }\href@noop {} {\bibfield
  {journal} {\bibinfo  {journal} {Physical Review A}\ }\textbf {\bibinfo
  {volume} {77}},\ \bibinfo {pages} {032322} (\bibinfo {year}
  {2008})}\BibitemShut {NoStop}%
\bibitem [{\citenamefont {Escand{\'o}n-Monardes}\ \emph
  {et~al.}(2024)\citenamefont {Escand{\'o}n-Monardes}, \citenamefont
  {Uzc{\'a}tegui}, \citenamefont {Rivera-Tapia}, \citenamefont {Walborn},\ and\
  \citenamefont {Delgado}}]{escandon2024estimation}%
  \BibitemOpen
  \bibfield  {author} {\bibinfo {author} {\bibfnamefont {J.}~\bibnamefont
  {Escand{\'o}n-Monardes}}, \bibinfo {author} {\bibfnamefont {D.}~\bibnamefont
  {Uzc{\'a}tegui}}, \bibinfo {author} {\bibfnamefont {M.}~\bibnamefont
  {Rivera-Tapia}}, \bibinfo {author} {\bibfnamefont {S.~P.}\ \bibnamefont
  {Walborn}},\ and\ \bibinfo {author} {\bibfnamefont {A.}~\bibnamefont
  {Delgado}},\ }\bibfield  {title} {\bibinfo {title} {Estimation of
  high-dimensional unitary transformations saturating the quantum
  cram{\'e}r-rao bound},\ }\href@noop {} {\bibfield  {journal} {\bibinfo
  {journal} {Quantum}\ }\textbf {\bibinfo {volume} {8}},\ \bibinfo {pages}
  {1405} (\bibinfo {year} {2024})}\BibitemShut {NoStop}%
\bibitem [{\citenamefont {Altepeter}\ \emph {et~al.}(2003)\citenamefont
  {Altepeter}, \citenamefont {Branning}, \citenamefont {Jeffrey}, \citenamefont
  {Wei}, \citenamefont {Kwiat}, \citenamefont {Thew}, \citenamefont
  {O’Brien}, \citenamefont {Nielsen},\ and\ \citenamefont
  {White}}]{altepeter2003ancilla}%
  \BibitemOpen
  \bibfield  {author} {\bibinfo {author} {\bibfnamefont {J.~B.}\ \bibnamefont
  {Altepeter}}, \bibinfo {author} {\bibfnamefont {D.}~\bibnamefont {Branning}},
  \bibinfo {author} {\bibfnamefont {E.}~\bibnamefont {Jeffrey}}, \bibinfo
  {author} {\bibfnamefont {T.}~\bibnamefont {Wei}}, \bibinfo {author}
  {\bibfnamefont {P.~G.}\ \bibnamefont {Kwiat}}, \bibinfo {author}
  {\bibfnamefont {R.~T.}\ \bibnamefont {Thew}}, \bibinfo {author}
  {\bibfnamefont {J.~L.}\ \bibnamefont {O’Brien}}, \bibinfo {author}
  {\bibfnamefont {M.~A.}\ \bibnamefont {Nielsen}},\ and\ \bibinfo {author}
  {\bibfnamefont {A.~G.}\ \bibnamefont {White}},\ }\bibfield  {title} {\bibinfo
  {title} {Ancilla-assisted quantum process tomography},\ }\href@noop {}
  {\bibfield  {journal} {\bibinfo  {journal} {Physical Review Letters}\
  }\textbf {\bibinfo {volume} {90}},\ \bibinfo {pages} {193601} (\bibinfo
  {year} {2003})}\BibitemShut {NoStop}%
\bibitem [{\citenamefont {Zou}\ \emph {et~al.}(1991)\citenamefont {Zou},
  \citenamefont {Wang},\ and\ \citenamefont {Mandel}}]{zou1991induced}%
  \BibitemOpen
  \bibfield  {author} {\bibinfo {author} {\bibfnamefont {X.}~\bibnamefont
  {Zou}}, \bibinfo {author} {\bibfnamefont {L.~J.}\ \bibnamefont {Wang}},\ and\
  \bibinfo {author} {\bibfnamefont {L.}~\bibnamefont {Mandel}},\ }\bibfield
  {title} {\bibinfo {title} {Induced coherence and indistinguishability in
  optical interference},\ }\href@noop {} {\bibfield  {journal} {\bibinfo
  {journal} {Physical Review Letters}\ }\textbf {\bibinfo {volume} {67}},\
  \bibinfo {pages} {318} (\bibinfo {year} {1991})}\BibitemShut {NoStop}%
\bibitem [{\citenamefont {Hochrainer}\ \emph {et~al.}(2022)\citenamefont
  {Hochrainer}, \citenamefont {Lahiri}, \citenamefont {Erhard}, \citenamefont
  {Krenn},\ and\ \citenamefont {Zeilinger}}]{hochrainer2022quantum}%
  \BibitemOpen
  \bibfield  {author} {\bibinfo {author} {\bibfnamefont {A.}~\bibnamefont
  {Hochrainer}}, \bibinfo {author} {\bibfnamefont {M.}~\bibnamefont {Lahiri}},
  \bibinfo {author} {\bibfnamefont {M.}~\bibnamefont {Erhard}}, \bibinfo
  {author} {\bibfnamefont {M.}~\bibnamefont {Krenn}},\ and\ \bibinfo {author}
  {\bibfnamefont {A.}~\bibnamefont {Zeilinger}},\ }\bibfield  {title} {\bibinfo
  {title} {Quantum indistinguishability by path identity and with undetected
  photons},\ }\href@noop {} {\bibfield  {journal} {\bibinfo  {journal} {Reviews
  of Modern Physics}\ }\textbf {\bibinfo {volume} {94}},\ \bibinfo {pages}
  {025007} (\bibinfo {year} {2022})}\BibitemShut {NoStop}%
\bibitem [{\citenamefont {Rajeev}\ and\ \citenamefont
  {Lahiri}(2026)}]{rajeev2026characterizing}%
  \BibitemOpen
  \bibfield  {author} {\bibinfo {author} {\bibfnamefont {S.}~\bibnamefont
  {Rajeev}}\ and\ \bibinfo {author} {\bibfnamefont {M.}~\bibnamefont
  {Lahiri}},\ }\bibfield  {title} {\bibinfo {title} {Characterizing a
  high-dimensional unitary transformation without measuring the qudit it
  transforms},\ }\href@noop {} {\bibfield  {journal} {\bibinfo  {journal}
  {Optics Letters}\ }\textbf {\bibinfo {volume} {51}},\ \bibinfo {pages} {853}
  (\bibinfo {year} {2026})}\BibitemShut {NoStop}%
\bibitem [{\citenamefont {Calvo}\ \emph {et~al.}(2006)\citenamefont {Calvo},
  \citenamefont {Pic{\'o}n},\ and\ \citenamefont {Bagan}}]{calvo2006quantum}%
  \BibitemOpen
  \bibfield  {author} {\bibinfo {author} {\bibfnamefont {G.~F.}\ \bibnamefont
  {Calvo}}, \bibinfo {author} {\bibfnamefont {A.}~\bibnamefont {Pic{\'o}n}},\
  and\ \bibinfo {author} {\bibfnamefont {E.}~\bibnamefont {Bagan}},\ }\bibfield
   {title} {\bibinfo {title} {Quantum field theory of photons with orbital
  angular momentum},\ }\href@noop {} {\bibfield  {journal} {\bibinfo  {journal}
  {Physical Review A}\ }\textbf {\bibinfo {volume} {73}},\ \bibinfo {pages}
  {013805} (\bibinfo {year} {2006})}\BibitemShut {NoStop}%
\bibitem [{\citenamefont {Plick}\ \emph {et~al.}(2013)\citenamefont {Plick},
  \citenamefont {Krenn}, \citenamefont {Fickler}, \citenamefont {Ramelow},\
  and\ \citenamefont {Zeilinger}}]{plick2013quantum}%
  \BibitemOpen
  \bibfield  {author} {\bibinfo {author} {\bibfnamefont {W.~N.}\ \bibnamefont
  {Plick}}, \bibinfo {author} {\bibfnamefont {M.}~\bibnamefont {Krenn}},
  \bibinfo {author} {\bibfnamefont {R.}~\bibnamefont {Fickler}}, \bibinfo
  {author} {\bibfnamefont {S.}~\bibnamefont {Ramelow}},\ and\ \bibinfo {author}
  {\bibfnamefont {A.}~\bibnamefont {Zeilinger}},\ }\bibfield  {title} {\bibinfo
  {title} {Quantum orbital angular momentum of elliptically symmetric light},\
  }\href@noop {} {\bibfield  {journal} {\bibinfo  {journal} {Physical Review
  A}\ }\textbf {\bibinfo {volume} {87}},\ \bibinfo {pages} {033806} (\bibinfo
  {year} {2013})}\BibitemShut {NoStop}%
\bibitem [{\citenamefont {Karimi}\ \emph {et~al.}(2014)\citenamefont {Karimi},
  \citenamefont {Boyd}, \citenamefont {De~La~Hoz}, \citenamefont {De~Guise},
  \citenamefont {{\v{R}}eh{\'a}{\v{c}}ek}, \citenamefont {Hradil},
  \citenamefont {Aiello}, \citenamefont {Leuchs},\ and\ \citenamefont
  {S{\'a}nchez-Soto}}]{karimi2014radial}%
  \BibitemOpen
  \bibfield  {author} {\bibinfo {author} {\bibfnamefont {E.}~\bibnamefont
  {Karimi}}, \bibinfo {author} {\bibfnamefont {R.~W.}\ \bibnamefont {Boyd}},
  \bibinfo {author} {\bibfnamefont {P.}~\bibnamefont {De~La~Hoz}}, \bibinfo
  {author} {\bibfnamefont {H.}~\bibnamefont {De~Guise}}, \bibinfo {author}
  {\bibfnamefont {J.}~\bibnamefont {{\v{R}}eh{\'a}{\v{c}}ek}}, \bibinfo
  {author} {\bibfnamefont {Z.}~\bibnamefont {Hradil}}, \bibinfo {author}
  {\bibfnamefont {A.}~\bibnamefont {Aiello}}, \bibinfo {author} {\bibfnamefont
  {G.}~\bibnamefont {Leuchs}},\ and\ \bibinfo {author} {\bibfnamefont {L.~L.}\
  \bibnamefont {S{\'a}nchez-Soto}},\ }\bibfield  {title} {\bibinfo {title}
  {Radial quantum number of laguerre-gauss modes},\ }\href@noop {} {\bibfield
  {journal} {\bibinfo  {journal} {Physical Review A}\ }\textbf {\bibinfo
  {volume} {89}},\ \bibinfo {pages} {063813} (\bibinfo {year}
  {2014})}\BibitemShut {NoStop}%
\bibitem [{\citenamefont {Mandel}\ and\ \citenamefont
  {Wolf}(1995)}]{mandel1995optical}%
  \BibitemOpen
  \bibfield  {author} {\bibinfo {author} {\bibfnamefont {L.}~\bibnamefont
  {Mandel}}\ and\ \bibinfo {author} {\bibfnamefont {E.}~\bibnamefont {Wolf}},\
  }\href@noop {} {\emph {\bibinfo {title} {Optical coherence and quantum
  optics}}}\ (\bibinfo  {publisher} {Cambridge University Press},\ \bibinfo
  {year} {1995})\BibitemShut {NoStop}%
\bibitem [{\citenamefont {Babazadeh}\ \emph {et~al.}(2017)\citenamefont
  {Babazadeh}, \citenamefont {Erhard}, \citenamefont {Wang}, \citenamefont
  {Malik}, \citenamefont {Nouroozi}, \citenamefont {Krenn},\ and\ \citenamefont
  {Zeilinger}}]{babazadeh2017high}%
  \BibitemOpen
  \bibfield  {author} {\bibinfo {author} {\bibfnamefont {A.}~\bibnamefont
  {Babazadeh}}, \bibinfo {author} {\bibfnamefont {M.}~\bibnamefont {Erhard}},
  \bibinfo {author} {\bibfnamefont {F.}~\bibnamefont {Wang}}, \bibinfo {author}
  {\bibfnamefont {M.}~\bibnamefont {Malik}}, \bibinfo {author} {\bibfnamefont
  {R.}~\bibnamefont {Nouroozi}}, \bibinfo {author} {\bibfnamefont
  {M.}~\bibnamefont {Krenn}},\ and\ \bibinfo {author} {\bibfnamefont
  {A.}~\bibnamefont {Zeilinger}},\ }\bibfield  {title} {\bibinfo {title}
  {High-dimensional single-photon quantum gates: concepts and experiments},\
  }\href@noop {} {\bibfield  {journal} {\bibinfo  {journal} {Physical Review
  Letters}\ }\textbf {\bibinfo {volume} {119}},\ \bibinfo {pages} {180510}
  (\bibinfo {year} {2017})}\BibitemShut {NoStop}%
\bibitem [{\citenamefont {Schlederer}\ \emph {et~al.}(2016)\citenamefont
  {Schlederer}, \citenamefont {Krenn}, \citenamefont {Fickler}, \citenamefont
  {Malik},\ and\ \citenamefont {Zeilinger}}]{schlederer2016cyclic}%
  \BibitemOpen
  \bibfield  {author} {\bibinfo {author} {\bibfnamefont {F.}~\bibnamefont
  {Schlederer}}, \bibinfo {author} {\bibfnamefont {M.}~\bibnamefont {Krenn}},
  \bibinfo {author} {\bibfnamefont {R.}~\bibnamefont {Fickler}}, \bibinfo
  {author} {\bibfnamefont {M.}~\bibnamefont {Malik}},\ and\ \bibinfo {author}
  {\bibfnamefont {A.}~\bibnamefont {Zeilinger}},\ }\bibfield  {title} {\bibinfo
  {title} {Cyclic transformation of orbital angular momentum modes},\
  }\href@noop {} {\bibfield  {journal} {\bibinfo  {journal} {New Journal of
  Physics}\ }\textbf {\bibinfo {volume} {18}},\ \bibinfo {pages} {043019}
  (\bibinfo {year} {2016})}\BibitemShut {NoStop}%
\bibitem [{\citenamefont {Brandt}\ \emph {et~al.}(2020)\citenamefont {Brandt},
  \citenamefont {Hiekkam{\"a}ki}, \citenamefont {Bouchard}, \citenamefont
  {Huber},\ and\ \citenamefont {Fickler}}]{brandt2020high}%
  \BibitemOpen
  \bibfield  {author} {\bibinfo {author} {\bibfnamefont {F.}~\bibnamefont
  {Brandt}}, \bibinfo {author} {\bibfnamefont {M.}~\bibnamefont
  {Hiekkam{\"a}ki}}, \bibinfo {author} {\bibfnamefont {F.}~\bibnamefont
  {Bouchard}}, \bibinfo {author} {\bibfnamefont {M.}~\bibnamefont {Huber}},\
  and\ \bibinfo {author} {\bibfnamefont {R.}~\bibnamefont {Fickler}},\
  }\bibfield  {title} {\bibinfo {title} {High-dimensional quantum gates using
  full-field spatial modes of photons},\ }\href@noop {} {\bibfield  {journal}
  {\bibinfo  {journal} {Optica}\ }\textbf {\bibinfo {volume} {7}},\ \bibinfo
  {pages} {98} (\bibinfo {year} {2020})}\BibitemShut {NoStop}%
\bibitem [{\citenamefont {Morizur}\ \emph {et~al.}(2010)\citenamefont
  {Morizur}, \citenamefont {Nicholls}, \citenamefont {Jian}, \citenamefont
  {Armstrong}, \citenamefont {Treps}, \citenamefont {Hage}, \citenamefont
  {Hsu}, \citenamefont {Bowen}, \citenamefont {Janousek},\ and\ \citenamefont
  {Bachor}}]{morizur2010programmable}%
  \BibitemOpen
  \bibfield  {author} {\bibinfo {author} {\bibfnamefont {J.-F.}\ \bibnamefont
  {Morizur}}, \bibinfo {author} {\bibfnamefont {L.}~\bibnamefont {Nicholls}},
  \bibinfo {author} {\bibfnamefont {P.}~\bibnamefont {Jian}}, \bibinfo {author}
  {\bibfnamefont {S.}~\bibnamefont {Armstrong}}, \bibinfo {author}
  {\bibfnamefont {N.}~\bibnamefont {Treps}}, \bibinfo {author} {\bibfnamefont
  {B.}~\bibnamefont {Hage}}, \bibinfo {author} {\bibfnamefont {M.}~\bibnamefont
  {Hsu}}, \bibinfo {author} {\bibfnamefont {W.}~\bibnamefont {Bowen}}, \bibinfo
  {author} {\bibfnamefont {J.}~\bibnamefont {Janousek}},\ and\ \bibinfo
  {author} {\bibfnamefont {H.-A.}\ \bibnamefont {Bachor}},\ }\bibfield  {title}
  {\bibinfo {title} {Programmable unitary spatial mode manipulation},\
  }\href@noop {} {\bibfield  {journal} {\bibinfo  {journal} {Journal of the
  Optical Society of America A}\ }\textbf {\bibinfo {volume} {27}},\ \bibinfo
  {pages} {2524} (\bibinfo {year} {2010})}\BibitemShut {NoStop}%
\bibitem [{Not()}]{Note-comp-form}%
  \BibitemOpen
  \href@noop {} {}\bibinfo {note} {The number of required choices can be
  reduced by considering more complex forms of the known unitary transformation
  as shown in Ref.~\cite{rajeev2026characterizing}.}\BibitemShut {Stop}%
\end{thebibliography}%

	
\end{document}